\documentclass[fleqn,usenatbib]{mnras}

\usepackage[T1]{fontenc}

\DeclareRobustCommand{\VAN}[3]{#2}
\let\VANthebibliography\thebibliography
\def\thebibliography{\DeclareRobustCommand{\VAN}[3]{##3}\VANthebibliography}

\usepackage{graphicx}	
\usepackage{amsmath}	
\usepackage{amssymb}	

\usepackage[normalem]{ulem}
\usepackage{newtxtext,newtxmath}

\newcommand{\msun}{\rm M_{\odot}}
\newcommand{\lsun}{\rm L_{\odot}}
\newcommand{\mstar}{\rm M_{\star}}
\newcommand{\lstar}{\rm L_{\star}}
\newcommand{\pc}{\mathrm{pc}}

\newcommand{\fid}{`Fiducial' }
\newcommand{\fidimf}{`Fiducial + IMF(Z)'}
\newcommand{\fidrt}{`Fiducial + RT' }
\newcommand{\fidrtimf}{`Fiducial + IMF(Z) + RT'}
\newcommand{\feh}{\rm [Fe/H]}
\newcommand{\alphafe}{\rm [O/Fe]}
\newcommand{\magv}{M_{\rm V}}
\newcommand{\rhalflight}{r_{\rm1/2}}
\newcommand{\mdm}{m_{\mathrm{DM}}}
\newcommand{\nsnii}{N_{\mathrm{SNeII}}}
\newcommand{\vdisp}{\sigma_{\star}}
\newcommand{\esnii}{\Tilde{E}_{\mathrm{SNeII}}}

\title[Varying IMF in ultra-faints]{EDGE: The sensitivity of ultra-faint dwarfs to a metallicity-dependent initial mass function}

\author[M. Prgomet et al.]{Mateo Prgomet,$^{1}$\thanks{E-mail: mateoprgomet@hotmail.com}
Martin P. Rey,$^{2, 1}$\thanks{E-mail: martin.rey@physics.ox.ac.uk}
Eric P. Andersson,$^{1}$
Alvaro Segovia Otero,$^{1}$
Oscar Agertz,$^{1}$
\newauthor
Florent Renaud,$^{1}$
Andrew Pontzen$^{3}$ and
Justin I. Read$^{4}$
\\
$^{1}$ Lund Observatory, Department of Astronomy and Theoretical Physics, Lund University, Box 43, SE-221 00 Lund, Sweden \\
$^{2}$ Sub-department of Astrophysics, University of Oxford, Keble Road, Oxford OX1 3RH, UK \\ 
$^{3}$ Department of Physics and Astronomy, University College London, London WC1E 6BT, UK\\
$^{4}$ Department of Physics, University of Surrey, Guildford GU2 7XH, UK \\
}

\date{Submitted to MNRAS}

\pubyear{2021}

\begin{document}
\label{firstpage}
\pagerange{\pageref{firstpage}--\pageref{lastpage}}
\maketitle

\begin{abstract}
Motivated by the observed bottom-light initial mass function (IMF) in faint dwarfs, we study how a metallicity-dependent IMF affects the feedback budget and observables of an ultra-faint dwarf galaxy. We model the evolution of a low-mass ($\approx 8 \, \times \, 10^{8} \,\msun$) dark matter halo with cosmological, zoomed hydrodynamical simulations capable of resolving individual supernovae explosions, which we complement with an empirically motivated subgrid prescription for systematic IMF variations. In this framework, at the low gas metallicities typical of faint dwarfs, the IMF of newborn stellar populations becomes top-heavy, increasing the efficiency of supernova and photoionization feedback in regulating star formation. This results in a 100-fold reduction of the final stellar mass of the dwarf compared to a canonical IMF, at fixed dynamical mass. The increase in the feedback budget is nonetheless met by increased metal production from more numerous massive stars, leading to nearly constant iron content at $z=0$. A metallicity-dependent IMF therefore provides a mechanism to produce low-mass ($\mstar \sim 10^3 \msun$), yet enriched ($\feh \approx -2$) field dwarf galaxies, thus opening a self-consistent avenue to populate the plateau in $\feh$ at the faintest end of the mass--metallicity relation.
\end{abstract}


\section{Introduction}

Improvements in deep, wide photometric imaging in recent years has allowed a rapid increase in the number of observed, faint dwarf galaxies (see \citealt{Simon2019} for a review), establishing and characterising a population of ultra-faint dwarf (UFD) galaxies ($\lsun \lesssim 10^{5} \, \lstar$). The low stellar masses of UFDs and their shallow potential wells make them particularly suited to constrain how star formation is regulated by feedback mechanisms within the interstellar medium (ISM; see \citealt{Somerville2015, Naab2017} for reviews). Achieving this, however, requires robust theoretical predictions, both resolving the ISM processes within such small galaxies while sampling their possible cosmological histories and environments. These imperatives place high demands on computing power, and has only recently been met by cosmological hydrodynamical simulations of galaxy formation (e.g. \citealt{Jeon2017, Maccio2017, Revaz2018, Munshi2019, Wheeler2019, Agertz2020EDGE, Applebaum2021}). 

Following this effort, discrepancies between individual predictions and the observed population of UFDs have become apparent. In particular, the faint-end of the mass--metallicity relation has proven a challenging and promising regime, due to its acute sensitivity to the modelling of star formation and feedback processes within a UFD (\citealt{Agertz2020EDGE}). In isolation, a dwarf's metallicity is set by (i) its overall metal production following stellar evolution and (ii) its ability to retain metals while regulating star formation in a shallow potential well. A detailed account of such ISM processes remains complex however, due to uncertainties in associating metal yields to massive stars at low metallicities (e.g. \citealt{Buck2021, Muley2021}) and to modelling the coupling between supernova and photoionization feedback to a dwarf's ISM (\citealt{Smith2019, Agertz2020EDGE, Smith2020PhotoRT}). These uncertainties propagate into inconsistent predictions for the chemical properties of field UFDs (see discussions in \citealt{Agertz2020EDGE,Applebaum2021}), with current models typically underenriching the faintest galaxies and struggling to explain the observed `plateau' of constant $\feh$ at the lowest stellar masses (\citealt{Kirby2013, Simon2019}).

Despite their sensitivity to internal ISM processes, these discrepancies might not be uniquely related to star formation and feedback within the dwarfs. The known population of UFDs is primarily observed within the virial radius of the Milky Way, thus also requiring a quantification of environmental processing (e.g. \citealt{Buck2019, Applebaum2021}). Tides from a massive host will preferentially strip the metal-poor outskirts, providing a mechanism to lower a dwarf's stellar mass at nearly constant $\feh$ and potentially populating the faint-end plateau. Alternatively, outflows from a nearby host might pollute and pre-enrich a dwarf's ISM at early times, leading to higher metallicities at a given stellar mass than expected in isolation. Although such environmental effects likely play a role in shaping the properties of Milky-Way satellites, their detailed contribution and importance remains obscured by their coupling with uncertainties in modelling astrophysical processes before infall (\citealt{Applebaum2021}).

Furthermore, observations of UFDs are operating at the limits of current instrumental capabilities -- deeper, more complete observations of the metal-poor outskirts of UFDs have revised and lowered determinations of dwarf metallicities (e.g. \citealt{Chiti2021}), providing a distinct way to ease tension with simulated objects. The next generation of instruments provides us with a unique opportunity to tell apart the respective roles of these mechanisms. Forthcoming wide and deep photometric surveys (e.g. the Vera Rubin Observatory, the Euclid and Nancy Roman Space Telescopes) are likely to start discovering isolated UFDs in the field (\citealt{Mutlu-Pakdil2021, Borlaff2022}). Establishing the metallicities and kinematics of such isolated objects, either through high-resolution spectroscopy of single faint stars (with e.g. the Thirty-Meter and Giant-Magellan Telescopes) or efficient scanning thanks to multiplexed spectrographs (e.g. \citealt{McConnachie2016}), will enable a clean separation on the respective roles of environmental and internal effects.

To further pursue model exploration and prepare for these prospects, we investigate in this work the sensitivity of dwarf metallicities to physical assumptions in computing the available feedback and metal budget, to which the initial mass function (IMF), i.e. the mass spectrum of a stellar population, is a key input. The IMF is generally observed to be universal within the solar neighbourhood (\citealt{Bastian2010}), although explaining this lack of variations with local interstellar conditions remains challenging (see \citealt{Kroupa2013} for a review). Extragalactic observations have in fact repeatedly suggested the existence of systematic IMF variations in massive elliptical galaxies (e.g. \citealt{vanDokkum2010, Cappellari2012, Ferreras2013, Martin-Navarro2015}, see \citealt{Smith2020IMFETGs} for a review) and in dwarf galaxies (\citealt{Dabringhausen2009, Dabringhausen2012, Geha2013, Gennaro2018}). They provide an empirical picture in which metallicity is a key driver of IMF variations (\citealt{Geha2013, Martin-Navarro2015, Gennaro2018, Liang2021}), suggesting an IMF becoming top heavy in metal-poor, faint dwarfs. 

None the less, there remains significant debate in the interpretation and significance of this correlation. Within faint dwarf galaxies and their old stellar populations, massive stars are long gone, thus only allowing a measurement of the low-to-intermediate-mass IMF slope. Inferring the high-mass slope can then be affected by finite observational depth (\citealt{El-Badry2017}, although see \citealt{Filion2020} for recent deeper efforts), and requires assuming a fixed universal, functional shape that might also vary (e.g. \citealt{Yan2020}). Furthermore, such observations are typically obtained from integrated observations across several stellar populations and dwarf galaxies. Theoretical models of the interstellar medium suggest metallicity to be a critical parameter for IMF variations due to its role in regulating the efficiency of cloud cooling and subsequent fragmentation (\citealt{Marks2012, Hopkins2013, Chabrier2014, Sharda2022BottomHeavyIMF}), but how to link, if at all, variations of the IMF over galactic scales to those on giant molecular cloud and stellar cluster scales, remains a challenge (e.g. \citealt{PflammAltenburg2009IMF, Kroupa2013, Dib2018, Jerabkova2018, Guszejnov2019}).

The acute sensitivity of UFDs to galaxy formation physics, thus, provides a chance to discriminate between models and clarify this picture -- at the low metallicities typical of UFDs ($\feh\approx-2$), a metallicity-varying IMF would strongly affect the metal and feedback budget from massive stars, motivating a quantification of its impact on observables and in particular the mass--metallicity relation. In this work, we thus extend ongoing efforts assessing how IMF variations affect galactic properties (e.g. \citealt{Guszejnov2017, Barber2018, Gutcke2019}) to the previously unexplored faintest galaxies. To this end, we perform cosmological, zoom simulations of a field low-mass galaxy as part of the EDGE project (Engineering Dwarfs at Galaxy formation’s Edge; \citealt{Rey2019UFDScatter, Rey2020, Rey2022EDGEHI}; \citealt{Agertz2020EDGE, Orkney2021}). Each simulation includes hydrodynamics, a comprehensive galaxy formation model (\citealt{Agertz2013,Agertz2020EDGE}; \citealt{Agertz2015}) and sufficient resolution ($\mdm = 960 \, \msun$, 3 pc) to resolve individual supernova explosions within the dwarf's ISM. We complement this setup with a new implementation of a metallicity-dependent IMF according to \citet{Geha2013} (see also \citealt{Kroupa2002} and \citealt{Marks2012}), which we describe in Section~\ref{METHOD}. We then show how a metallicity-dependent IMF boosts both feedback and metal production in the faintest dwarfs, leading to a reduction in their final stellar masses while maintaining significant iron enrichment (Section~\ref{RESULTS}). We conclude in Section~\ref{CONCLUSIONS}.

\section{Methods} \label{METHOD} 

As we aim to quantify and isolate the impact of a varying IMF, we focus on a single UFD with a final dynamical halo mass of $\approx 8 \, \times \, 10^{8} \ \msun$. The host dark matter halo evolved with our baseline galaxy formation model (hereafter `Fiducial') was first presented in \citet{Agertz2020EDGE}. We briefly summarize the main features of the `Fiducial' model in Section~\ref{sec:Agertz2020}, before describing how we complement it with an implementation of IMF variations in Section~\ref{sec:imfmethods}.

\subsection{Fiducial numerical setup} \label{sec:Agertz2020}

We target an isolated dark matter halo (no neighbours more massive than it within five virial radii) embedded within a cosmic underdensity to minimize the chances of environmental interactions (see \citealt{Agertz2020EDGE} for further details and \citealt{Orkney2021} for a visual). We construct zoomed, cosmological initial conditions for this dark matter halo using the \textsc{genetIC} software (\citealt{Stopyra2021}) with dark matter particles of mass $\mdm = 945 \, \msun$ and baryonic resolution $\Omega_{\mathrm{b}} / \Omega_{\mathrm{m}} \, \mdm = 161 \, \msun $. We follow the evolution of stars, dark matter and gas using the adaptive-mesh-refinement code \textsc{ramses} (\citealt{Teyssier2002}), achieving a maximum spatial resolution of 3 pc across the dwarf's ISM. We account for the cooling of gas out of equilibrium (\citealt{Rosdahl2013}), and the formation of stars using a Schmidt-like law with a star formation efficiency of $10\%$ (\citealt{Agertz2020EDGE}). Star particles are formed through stochastic sampling on a cell-by-cell basis at every simulation fine step (\citealt{Rasera2006}) with a birth mass of $300\, \msun$, sufficient to ensure a complete sampling of the canonical IMF in each stellar population (e.g. \citealt{Smith2021IMF}). We note that complete sampling for each individual population is not guaranteed once the IMF is allowed to vary (Section~\ref{sec:imfmethods}), but we show in Appendix~\ref{app:imfsampling} that the IMF averaged over the multiple stellar populations of a UFD's star formation history is well-sampled, thus ensuring the robustness of the total, integrated feedback and metal production budgets. As we wish to compare results with our \fid model, we keep this parameter fixed in this study, and plan to explore in future work the importance of IMF sampling (see also discussion in Section~\ref{CONCLUSIONS}).

We track the injection of momentum, energy, iron, and oxygen from stellar winds, Type II and Type Ia supernovae in the dwarf's ISM (SNe II, SNe Ia, see \citealt{Agertz2013, Agertz2015, Agertz2020EDGE} for details). Our resolution allows us to directly inject thermal energy associated with supernovae and self-consistently follow the build-up of momentum by solving the hydrodynamics equations, thus, strongly reducing uncertainties associated with modelling supernova feedback (\citealt{Kim2015, Martizzi2015, Smith2019}). In addition, some of our simulations model photoionization feedback with radiative transfer, following the evolution of radiation discretized into six photon groups to form a low-resolution spectrum, and computing its advection across cell boundaries using \textsc{ramses-rt} (\citealt{Rosdahl2013,Rosdahl2015RAMSESRT}, see \citealt{Agertz2020EDGE} for further information). We record the evolution of two metals, iron and oxygen, and we initialize oxygen at $10^{-3} Z_{\odot}$ to account for unresolved enrichment from Population-III stars. We model reionization as a homogeneous time-dependent, ultraviolet (UV) background as in \citet{Faucher-Giguere2009}.

\subsection{Implementing a metallicity-dependent initial mass function} \label{sec:imfmethods}

We complement the \fid model with an IMF which varies depending on metallicity. Each stellar particle is treated as a single stellar population (SSP) with an associated two-part Kroupa IMF (\citealt{Kroupa2001}):
 \begin{equation} \label{IMF}
    \Phi(M)
    =
    A
    \begin{cases}
        \begin{array}{lll}
        C_1 M^{-1.3}   & \mathrm{for} & 0.1 \leq M < 0.5 \ \msun \, ,\\
        C_2 M^{-\alpha} & \mathrm{for} & 0.5 \leq M < 100 \ \msun \, , \\
        \end{array}
    \end{cases}
\end{equation}
where $\Phi$ is the number of stars of mass $M$, $A$ is a normalization constant and $C_1, C_2$ ensure continuity between the two mass regimes. For our \fid model with a fixed IMF, the high-mass slope is set to $\alpha = 2.35$. 

In this work, we assume that the IMF varies with metallicity using an observationally motivated relation (\citealt{Geha2013}, see also \citealt{Kroupa2002}), given as 
\begin{equation}\label{Variation}
    \alpha = 
        \begin{cases}
        0.5 \times \mathrm{[Fe/H]}_i + 2.35 & \text{if $\mathrm{[Fe/H]}_{i} > -3$} \, , \\
        0.85 & \text{if $\mathrm{[Fe/H]}_{i} \leq -3$} \, , \\
        \end{cases} 
\end{equation}
where ${\feh}_i$ is the metallicity of star particle $i$. Equation~\ref{Variation} follows an empirical determination of IMF variations in UFDs (\citealt{Geha2013}, figure 5), which we extrapolate from the last observational point at ${\feh}=-2.5$ to ${\feh}=-3.0$\footnote{We impose a floor to the extrapolation at the lowest metallicities, to prevent extreme stellar populations in the first generation of stars initialized with vanishing ${\feh}$. We verified that such events represent at most $\sim 3$ per cent of the total number of stellar populations, following rapid enrichment of the ISM to ${\feh}\geq-3$ by the first few SNe II.} and renormalize to match our fiducial IMF (\citealt{Kroupa2001}) at solar metallicity. We stress that IMF variations driven through ISM metallicity remain theoretically motivated (e.g. \citealt{Sharda2022BottomHeavyIMF}) although debated, suggesting future implementations of additional variation channels through molecular cloud density (e.g. \citealt{Marks2012}) or ISM turbulence (e.g. \citealt{Hopkins2013}) are motivated. 

A direct consequence of Equation~\ref{Variation} is that the number of massive stars in each stellar population increases as metallicity decreases ($\alpha$ decreases and the IMF becomes more top heavy). Quantitatively, the mass fraction of massive stars ($M > 8 \, \msun$) at solar metallicity is 19 per cent, compared to 94 per cent at ${\feh}=-3$. Modifying the number of massive stars affects the energy, metal and radiation budget available in each stellar population, which we quantify now.

\begin{figure}
    \centering
    \includegraphics[width=\columnwidth]{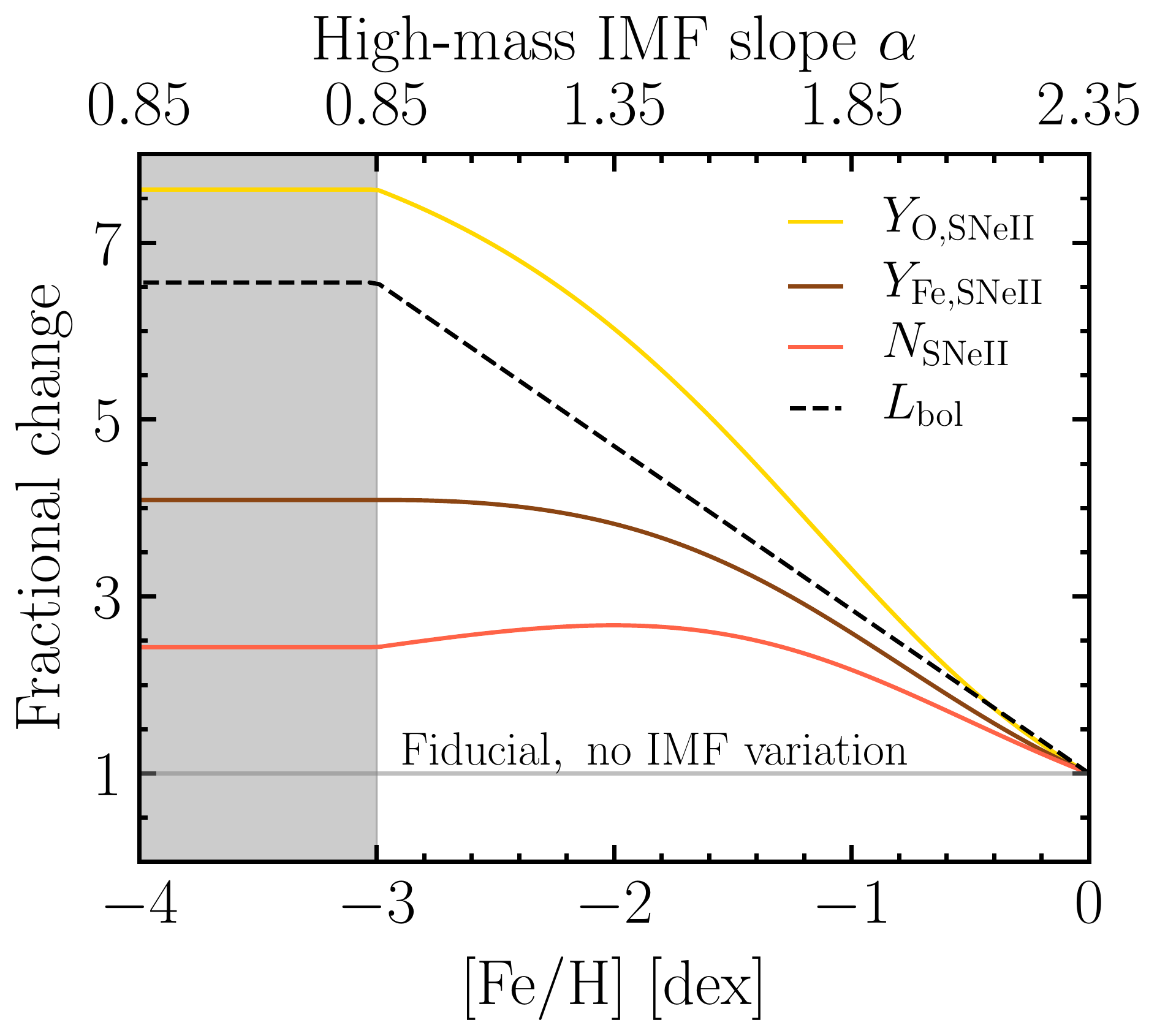}

    \caption{Modifications to the feedback and metal budget for a stellar population with an IMF varying with metallicity. As the high-mass slope decreases (top axis) with decreasing metallicities (bottom axis), the IMF becomes more top-heavy, boosting the number of available SNe II (red) and bolometric luminosity (black dashed). In parallel, oxygen (yellow) and iron (brown) production is enhanced due to the increased number of massive stars. We cap the metallicity evolution for the most metal-poor populations ($\feh \leq -3$, grey shading).}
    \label{fig1}
\end{figure}

Figure~\ref{fig1} shows the fractional change in the number of SNe II in each stellar population, $\nsnii$ (red), compared to a fiducial, fixed IMF (grey). In the range most relevant to UFD galaxies ($ -3 \leq{\feh}\leq-1$, which encompasses the complete metallicity distribution within our simulated galaxies), the number of SNe II is  boosted by a roughly constant factor of $2.5$ compared to the fiducial case. This directly translates into an increased energy budget for feedback through $E_{\rm SNII} = \nsnii \, \esnii$ as, following our fiducial model, we assume all individual supernovae events to release the same energy $\esnii = 10^{51} \rm \, erg$. We leave an exploration of progenitor-dependent explosion budgets (e.g. \citealt{Sukhbold2016}) as future work.

We note a peak, followed by a slight lessening, in the increase of $\nsnii$ at the lowest metallicities. This feature is due to our assumption that SNe II are sampled from progenitors with masses ranging from 8 to $40 \, \msun$ within their parent stellar particle -- a specific IMF slope hence maximizes the number of massive stars within this mass range (here $\alpha\approx1.45$ at $\feh \approx -1.8$). A more top-heavy IMF then overproduces supermassive stars that are not accounted as SNe II. Given the remaining uncertainties in the mass upper limit for progenitors of SNe II (see e.g. \citealt{Janka2012, Janka2016} and references therein), we leave a detailed quantification of its importance in determining the feedback budget of a dwarf to a future study. 

We further update the metal budget by computing progenitor-mass dependent yields for each SNe II (\citealt{Kim2014}, equations 5 and 6) and show the total mass of iron (brown) and oxygen (yellow) released by SNe II in Figure~\ref{fig1} as the IMF is varying. Following the increase in $\nsnii$, the overall metal production is boosted compared to the fiducial, baseline scenario. Progenitor-mass-dependent yields, however, introduce different weightings for each individual metal as SNe II originating from massive progenitors contribute exponentially more than low-mass events. Oxygen production is thus boosted by up to a factor $\approx 7.5$, while iron production can be increased up to a factor $\approx 4$.

Finally, we account for modifications to the photon budget introduced by a metallicity-dependent IMF, as radiative feedback is a key process to regulating star formation in UFDs (\citealt{Agertz2020EDGE, Smith2020PhotoRT}). To capture the early, photoionizing phase during which the total luminosity of each SSP is UV-dominated, we rescale uniformly the flux in each fiducial photon group for $\sim 40$ Myr -- the time after which all massive stars tracked in the model have turned supernova. This is necessary to ensure the correct bolometric, and hence UV, luminosity for each SSP. We compute a library of IMF-weighted, bolometric luminosities for zero-age $300\, \msun$ SSPs using \textsc{starburst99} (SB99; \citealt{Leitherer1999}), scanning through the IMFs slopes expected for ISM metallicities up to solar in our model.\footnote{Although \textsc{starburst99} is only calibrated down to $0.05 \, Z_{\odot}$, the metallicity dependency of stellar evolution in this regime is subdominant compared to changes to the number of massive stars as we vary the IMF.} For each IMF mapping onto a metallicity, we then compute the boost in luminosity compared to the canonical slope, which we linearly interpolate over the metallicity range to obtain the uniform rescaling of photon groups (black dashed in Figure~\ref{fig1}). As expected from an increased number of massive stars with more top-heavy IMFs, the integrated bolometric luminosity of each SSP increases with decreasing metallicity, by up to a factor of $\sim6.5$ compared to the fiducial case.

We stress that modifying the bolometric luminosity only provides a first approximation to the available photon budget with top-heavy IMFs -- in principle, each photon group should be rescaled independently to account for changes in the spectral energy distribution. Our model is nonetheless a lower limit for the UV flux, as top-heavier IMFs would shift the peak of the SSP spectrum towards shorter wavelengths, further increasing the energy of photoionization feedback. We leave an implementation of such schemes as future work.

We identify the dark matter halo of galaxies using the \textsc{hop} halo finder (\citealt{Eisenstein1998}) and extract stellar properties using the \textsc{pynbody} (\citealt{Pontzen2013}) library. We derive stellar spectra for each IMF associated with a stellar particle using SB99, and perform mock observations of the galaxy by computing the radiative transfer through astronomical dust using a method similar to the \textsc{sunrise} code (\citealt{Jonsson2006}) and a dust attenuation curve from \citet{Li2001}. We pick a random line of sight to compute surface brightness radial profiles and determine the half-light radius, $\rhalflight$, of each simulated galaxy. Since observational abundances are determined from bright stars in the inner galaxy (e.g. \citealt{Simon2007, Kirby2010}; S. Kim et al., in preperation), we select all stars within $2 \, \rhalflight$ to determine the galaxy's mean iron metallicity, $\feh$ (\citealt{Agertz2020EDGE}, eq. 5), the 1D equivalent of velocity dispersion, $\vdisp$ and total absolute V-band magnitude, $\magv$ ({\citealt{Agertz2020EDGE}, sec. 3.4).

\begin{figure}
    \centering
    \includegraphics[width=\columnwidth]{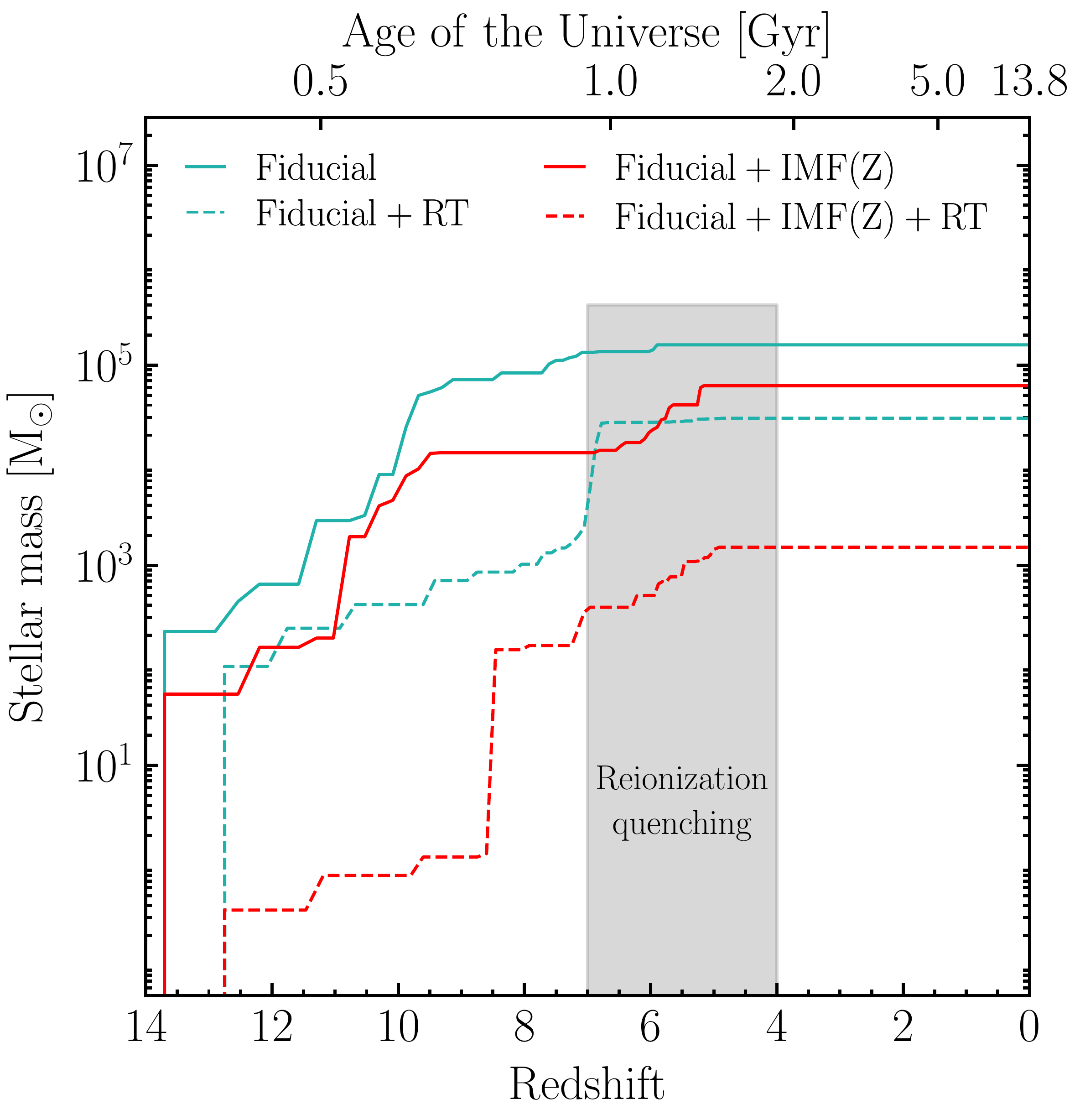}
    
    \caption{Stellar mass growth of each UFD as we resimulate their same dark matter host halo with varying physical models. Allowing the IMF to vary with metallicity (solid red) enhances the number of SNe II (Figure~\ref{fig1}), causing a reduction in the stellar mass of the dwarf compared to the fiducial scenario (solid turquoise). This effect combines with the increased effectiveness of photoionization feedback under a varying IMF (dashed red), further reducing the final stellar mass of our ultra-faint by a dex compared to the fiducial scenario accounting for radiative feedback (dashed turquoise).} 
    \label{fig2}
\end{figure}

\section{Results} \label{RESULTS}
Having accounted for variations in energy, metal, and photon budget with metallicity, we are now equipped to assess how IMF variations affect the observable properties of an UFD galaxy.

\subsection{Stellar masses and sizes with a varying IMF} \label{RESULTS:masses}

\begin{figure}
    \centering
    \includegraphics[width=\columnwidth]{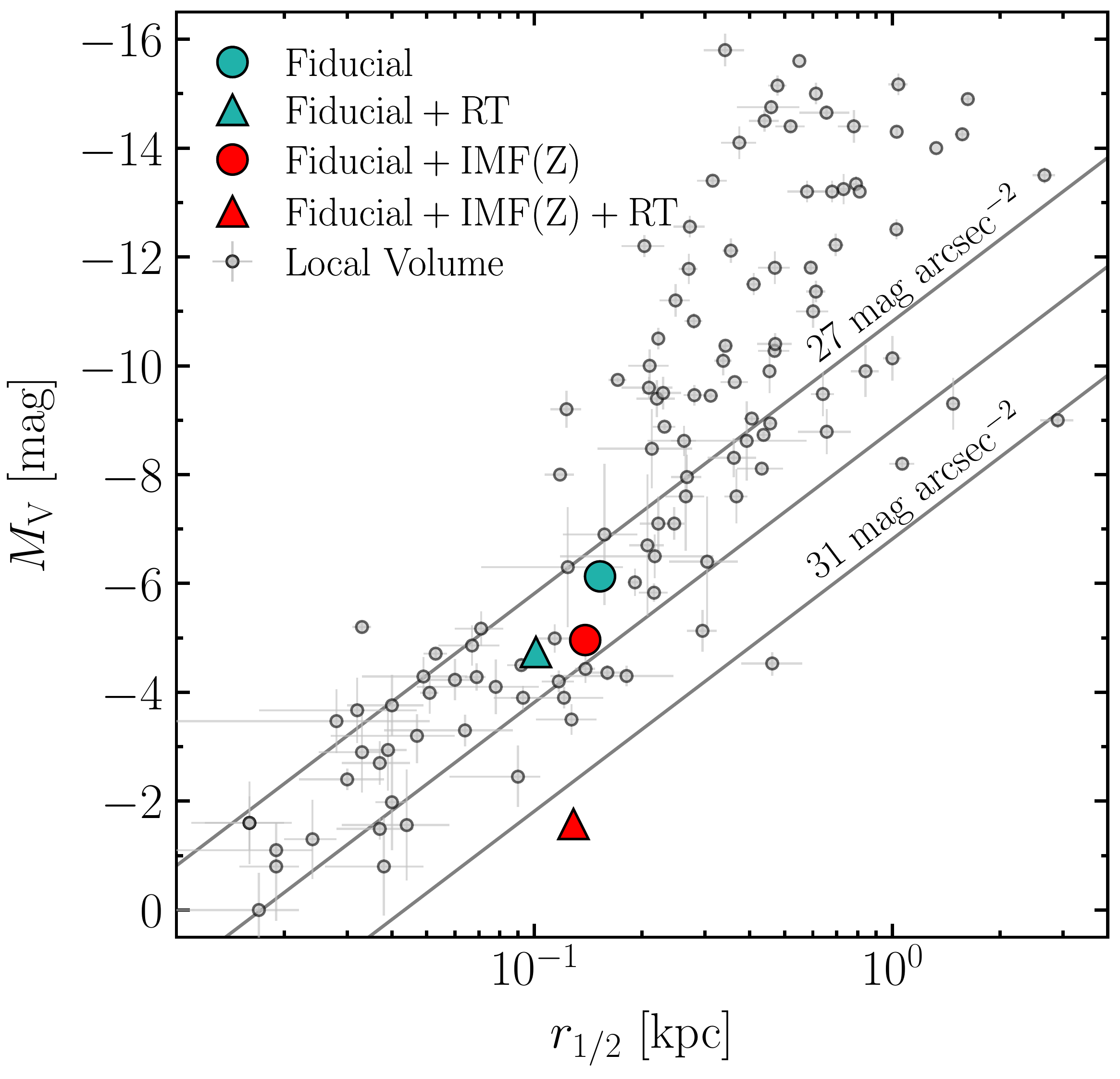}
    \caption{Absolute, total V-band magnitudes and half-light radii of our simulated galaxies compared to a compilation of observed, Local Group dwarf galaxies (grey points). As the budget of feedback increases through a metallicity-dependent IMF (turquoise to red), the total magnitude of our simulated ultra-faint increases, mirroring its stellar mass (Figure~\ref{fig2}). All galaxies, however, have similar sizes ($\rhalflight \sim 150 \, \pc$), yielding both overall fainter and more diffuse dwarfs (diagonal grey lines show constant central surface brightnesses). Our most extensive model including a varying IMF and increased radiative feedback (red triangle) lacks observed counterparts -- its properties place it below the detection capabilities of current instruments but future photometric surveys (e.g. the Vera Rubin Observatory, the Euclid and Nancy Roman Space Telescopes) will enable us to verify its physical nature.}
    \label{fig3}
\end{figure}

In Figure~\ref{fig2}, we compare the stellar mass assembly of our simulated UFDs, evolved with and without a metallicity-dependent IMF. We select all stars at $z=0$ encompassed within a sphere defined by the virial radius and centered around the galaxy, extract their birth times and accumulate their final masses over time.\footnote{The final masses of each star particle are, thus, computed after their lifetime's mass-loss from supernovae and winds, which can lead to total galactic stellar masses lower than an individual birth mass early-on.} This provides an archaeological record of the stellar mass across all progenitors, as would be observed from a dwarf in the Local Group today.

We first note that star formation proceeds up to $z\sim4$ in all simulated versions of our UFD and is quenched permanently after this time. This quenching is imposed by cosmic reionization which, by $z\sim 8-6$, has heated the intergalactic medium up to $T\sim 10^{4}~\rm K$. This increased pressure support prevents gas accretion into the potential well of small dwarf galaxies (final dynamical masses $\lesssim 10^9\, \msun$, e.g. \citealt{Efstathiou1992, Hoeft2006, Okamoto2008, Noh2014}). Self-shielded gas within the galaxy can allow star formation to proceed up to $z\sim4$ (\citealt{Susa2004, Onorbe2015, Agertz2020EDGE}), but the gas reservoir of the galaxy is never replenished, eventually leading to the definitive cessation of star formation. 

We now compare the UFD evolved with the \fid model detailed in \citet{Agertz2020EDGE} to our new implementation allowing for a metallicity-dependent IMF (\fidimf). A varying IMF leads to systematically lower stellar masses at all times, and a reduction in the final stellar mass by a factor of $\sim2.5$ compared to the \fid case. This reduction is naturally explained by the increased number of massive stars with a metallicity-varying IMF -- all stellar particles within our ultra-faint have $\feh \leq -1.1$ at $z=0$, and hence see their number of SNe II at least doubled (Figure~\ref{fig1}).

Turning to photoionisation feedback and its combined effect with a varying IMF, we first introduce the \fidrt version of the UFD from \citet{Agertz2020EDGE}, recovering that accounting for additional radiative feedback from massive stars lowers the final stellar mass of the galaxy by a factor of $\approx 5.4$ compared to the \fid model. Further accounting for the increased photoionizing budget from a varying IMF (recall Figure~\ref{fig1}), here denoted \fidrtimf, results in a much stronger suppression of up to 2 dex compared to the \fid model. By increasing the available supernova and radiative feedback budget, a metallicity-varying IMF strongly reduces star formation in UFDs before cosmic reionization permanently shuts down their gas supply. It therefore predicts the existence of low mass, field ultra-faints, here with $\mstar \sim 10^{3} \, \msun$ and dynamical masses $\approx 8 \, \times \, 10^{8} \, \msun$ at $z=0$.

Figure~\ref{fig3} then compares the sizes and total absolute V-band magnitudes of our simulated objects to a compilation of known, satellite dwarf galaxies in the Local Group (grey points; \citealt{McConnachie2012, Kirby2014, Homma2019, Torrealba2019, Simon2019}). As expected from the evolution of the final stellar masses, the total V-band magnitude increases as we account for a varying IMF and radiative feedback. All models, however, produce an UFD with a similar half-light radius of $\rhalflight \approx 150-200 \, \pc$, and we further verified similar 1D stellar velocity dispersions ($\vdisp \approx 5-6 \, \rm km/s$). We therefore reaffirm that re-simulating the same dark matter history with variations to the feedback budget yield similar dynamical properties for UFDs (\citealt{Agertz2020EDGE}), as these are primarily set by the mass and growth history of the host dark matter halo (e.g. \citealt{Rey2019UFDScatter}).

A direct consequence of reducing stellar mass at fixed size is to create both overall fainter, and lower surface brightness dwarfs (constant central surface brightness lines are shown in grey in Figure~\ref{fig3}). All simulations are compatible with observed systems, except for the model featuring both IMF variations and radiative transfer (red triangle), which creates a more diffuse ultra-faint than currently observed. 

This mismatch is best explained by the current observational difficulties of detecting objects at such low surface brightnesses. Given its total magnitude and size, this simulated ultra-faint would currently be undetectable within the virial radius of the Milky Way by a modern imaging survey (e.g. DES; \citealt{Drlica-Wagner2020}, figure 5), let alone in the field where our simulated objects reside. Another source of uncertainty in this comparison lies in that observed faint dwarfs all reside within the Milky Way. Their sizes and stellar contents could thus potentially be processed by their host's tidal environment during infall. However, very few, if any, UFDs appear to be tidally limited today based on their orbits (\citealt{Simon2018}), their stellar kinematics lack tidal processing signatures (\citealt{Zoutendijk2021TidesAndCores}), while the identification of member stars out to large radii in at least one case further argues against significant tidal stripping (\citealt{Chiti2021}). It, thus, remains unclear whether, and by how much, tidal processing affects the observables of currently known UFDs, suggesting our comparison is likely to be valid to leading order. Furthermore, dwarf galaxy sizes could also be underestimated at the faintest end, due to finite observational depth in the outskirts of galaxies (e.g \citealt{Chiti2021}). We plan in future work to quantify the importance of this effect when comparing simulated and observed objects (S. Kim et al., in preperation).

Such faint and diffuse systems beyond observational reach have been predicted to arise through different physical mechanisms, such as tidal stripping in a Milky-Way environment in which observed systems reside (e.g. \citealt{Applebaum2021}), from an altogether different modelling of feedback and galaxy formation physics (e.g. \citealt{Wheeler2019}), or from the diversity of possible assembly histories in field UFDs (\citealt{Rey2019UFDScatter}) -- a metallicity-dependent IMF therefore provides a distinct and new avenue to create faint UFDs beyond the reach of current surveys. 

Our results thus suggest IMF-variations as a plausible framework to interpret photometric observations of UFDs, although it remains to be quantified whether systematic variations across every dwarf's ISM would be consistent with the overall observed population. Such quantification requires us to both (i) sample multiple cosmological histories at a given halo mass to quantify how IMF variations affect the scatter in stellar masses and sizes (see also \citealt{Rey2019UFDScatter}), and (ii) sample alternate halo masses to pinpoint the response of observables as the IMF tends towards its canonical shape at higher masses and metallicities. Our study provides a strong motivation for future studies to pursue this assessment, in particular to establish which levels of IMF variations are compatible with the more complete census of faint dwarfs' sizes and absolute magnitudes that will be obtained by upcoming photometric surveys such as the Vera Rubin Observatory and Nancy Roman telescope. We leave this to future work and focus now on the imprint of IMF variations on spectroscopic properties of faint dwarfs.

\subsection{Iron enrichment and the mass--metallicity relation} \label{RESULTS:metallicity}

\begin{figure}
    \centering
    \includegraphics[width=\columnwidth]{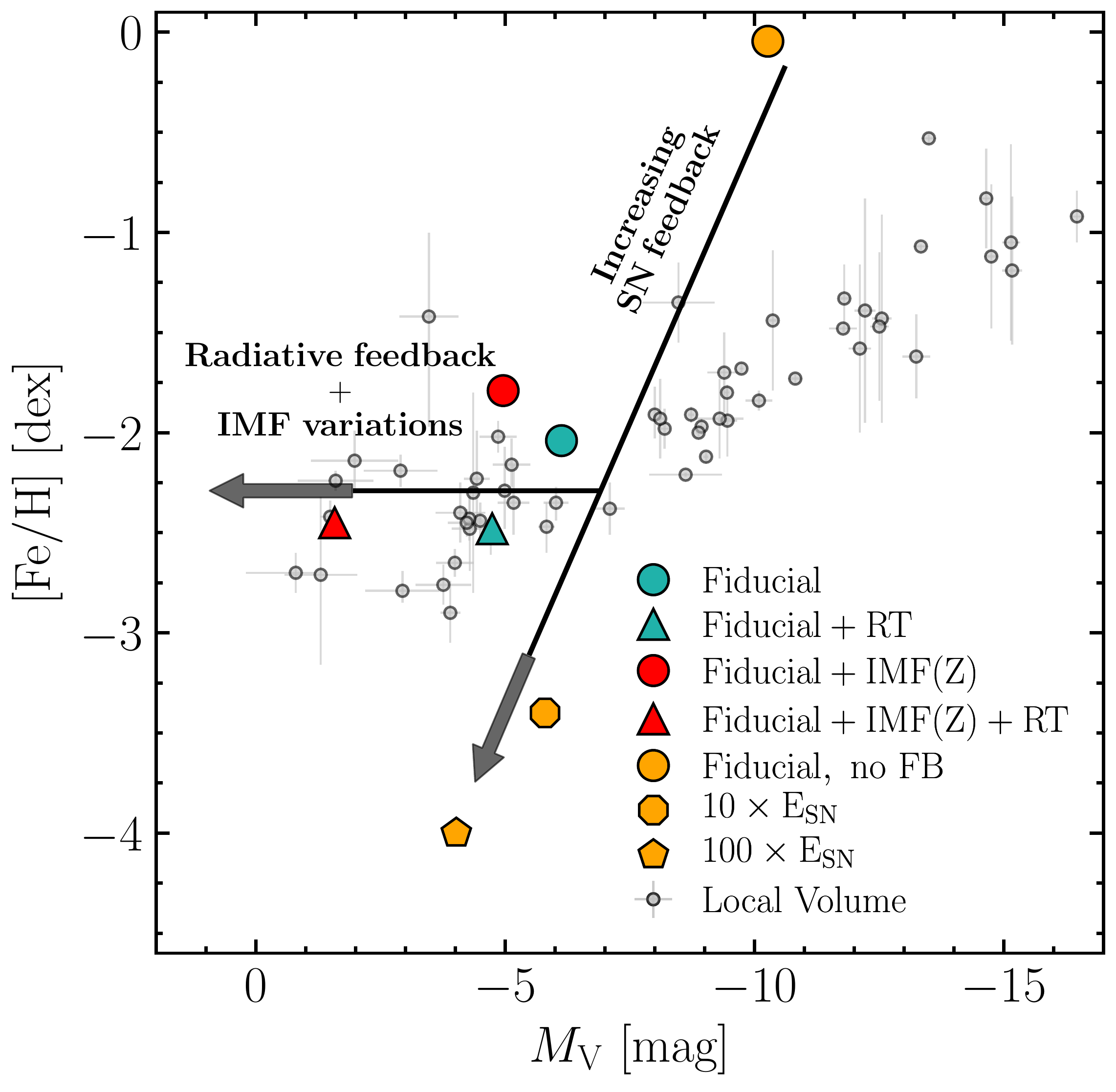}

    \caption{Mean iron metallicities and absolute V-band magnitudes of our simulated galaxies. Compared to the fiducial model (turquoise circle), radiative feedback (turquoise triangle) regulates star formation less explosively, lowering the final stellar mass while retaining similar metal amounts. Allowing for IMF variations (red circle) yields comparable outcomes, as the increased number of supernovae is compensated by an increased metal production (Figure~\ref{fig1}). Combining these two effects (red triangle) amplifies this trend, leading to a faint but significantly enriched field ultra-faint within the observed metallicity `plateau' from Local Group galaxies (both confirmed and candidates reported as grey points). Enhancing the strength of supernova feedback (yellow) also produces fainter systems but drastically reduces their mean metallicities, discrepant with observations due to overly ejective regulation of star formation.}
    \label{fig4}
\end{figure}

The mass--metallicity relation is a sensitive probe of alterations to the feedback, outflow and metal budgets in a dwarf galaxy (\citealt{Agertz2020EDGE}), all of which are modified when introducing a metallicity-dependent IMF (Figure~\ref{fig1}, see also e.g. \citealt{Koeppen2007}). We quantify the impact of a varying IMF on the metallicity content of our simulated dwarfs by showing the mass weighted, average, iron metallicity as a function of their absolute, total V-band magnitude in Figure~\ref{fig4}.

We start by comparing the impact of IMF variations without accounting for photoionization feedback from massive stars (circles). As discussed in Section~\ref{RESULTS:masses}, allowing for a metallicity-dependent IMF increases the number of SNe II, reducing the galaxy's final stellar mass and making it fainter overall. However, the overall production of iron by a stellar population is also significantly boosted when allowing for a varying IMF, resulting in an increased \feh \ from $-2.1$ to $-1.8$. A metallicity-dependent IMF therefore provides a way to decrease an ultra-faint's stellar mass, while maintaining a similar metal content within the galaxy by increasing metal production.

In this first scenario, star formation is regulated by repeated blowouts following the explosions of supernovae in the dwarf's ISM. Photoionization from massive stars, however, provides an alternative regulation mechanism, keeping the dwarf's ISM warmer, decreasing the clustering of supernovae and in turn the strength of galactic outflows (\citealt{Agertz2020EDGE, Smith2020PhotoRT}). This allows for greater metal retention within the dwarf galaxy, thus also resulting in a lowering of the galaxy's stellar mass at roughly fixed \feh\ (turquoise triangle).

These two effects combine further when accounting for both photoionization feedback and a metallicity-dependent IMF (red triangle). The overproduction of massive stars increases the effectiveness of radiative feedback by boosting the early UV luminosity of each stellar population (Figure~\ref{fig1}). This results in an ever-fainter dwarf galaxy, but with a metallicity only decreased from --2.1 to --2.5 compared with the fiducial case. Thus, an increasingly top-heavy IMF in metal-poor ISMs lowers the stellar masses of UFDs by up to 2 dex, without significantly lowering their metal content by increasing both metal production per population and retention within the ISM.

The significance of this statement is emphasized by comparing our simulated galaxies with an observed sample of dwarf galaxies around our Milky Way (grey points; \citealt{Kirby2013, Simon2019} showing both candidates and confirmed UFDs). An apparent feature in the observed mass--metallicity relation is the plateau in metallicity at $\feh \approx -2.5$ for faint dwarf galaxies with $\magv \gtrsim -7.5$. This plateau has proven to be a significant predictive challenge for current numerical simulations of UFDs (\citealt{Agertz2020EDGE}), with most studies underenriching dwarf galaxies in iron at the faintest end both when modelling faint dwarfs in the field and within a Milky-Way environment (\citealt{Maccio2017, Wheeler2019, Applebaum2021}). The robustness of the faint-end plateau remains uncertain due to the small number of confirmed observations (\citealt{Simon2019}), but its ability to discriminate between galaxy formation models makes it a particularly promising prospect in the near future.

We further show how such enrichment patterns cannot be reproduced by remaining uncertainties in supernova feedback. Although our resolution allows us to accurately follow the explosion of individual supernovae, pinpointing their clustering and subsequent coupling to the dwarf's ISM remains challenging (\citealt{Smith2019, Agertz2020EDGE, Smith2020PhotoRT}). We show re-simulations of the \fid galaxy artificially increasing the strength of each supernovae event 10-fold, 100-fold and turning off all feedback channels (yellow symbols, see also \citealt{Agertz2020EDGE}). As the strength of supernova feedback increases, blowouts of gas become more efficient at regulating star formation, reducing the final stellar mass of the dwarf. This more ejective regulation mode, however, steeply lowers the retained metal content, resulting in underenriched dwarfs at a given stellar mass.

Our study therefore reaffirms the faint end of the mass--metallicity relation as an acute test of galaxy formation and ISM physics. We demonstrate that observationally motivated IMF variations (\citealt{Geha2013}) provide a self-consistent mechanism to enrich isolated, field UFD galaxies to the observed metallicity `plateau', strongly motivating further assessments of the role played by IMF variations in shaping the observables of the dwarf galaxy population.

In particular, quantifying whether IMF variations as considered in this work would stay consistent with the observed mass--metallicity relation at higher masses should be a key focus for future studies. As stellar masses and ISM metallicities increase, the IMF will smoothly transition towards its canonical value at solar metallicity to remain consistent with Milky-Way determinations. This potentially provides a natural explanation for the transition from a plateau to a slope in the mass--metallicity relation, but might also over-enrich dwarfs at higher masses compared to current data. Pinpointing which levels of IMF variations are allowed across the classical and faint dwarf population will thus be key to constrain our understanding of star formation. Another interesting signature to differentiate a varying from a canonical IMF could be to target the significant overproduction of $\alpha$ elements due to IMF variations (recall Figure~\ref{fig1}, gold). As an example, the average $\alphafe$ abundance is increased from 0.75 to 0.92 when complementing the \fidrt model with IMF variations. Similar $\alpha$--enrichment patterns have been reported in local UFDs of similar masses (e.g. \citealt{Vargas2013, Frebel2014, Simon2019}), but abundance data is limited to a handful of bright stars within each observed dwarf and the galaxy-to-galaxy and star-to-star scatter makes quantitative comparisons currently challenging.

Finally, implementing a varying IMF at these low metallicities produces rather extreme stellar populations, some only consisting of a few massive stars. We show in Appendix~\ref{app:imfsampling} that the integrated metal and feedback budget over a dwarf's history is robust to the mass and phase--space sampling of the IMF, but the impact of sampling noise on the instantaneous coupling between feedback and the local ISM remains to be quantified with strongly top-heavy IMFs (see also \citealt{Kroupa2013, Su2018, Grudic2019, Applebaum2020, Smith2021IMF} for discussions with a canonical IMF). In future work, we plan on coupling our varying-IMF implementation with the ability to resolve energy and metal injection from individual massive stars in a galaxy's ISM (e.g. \citealt{Emerick2019, Andersson2020, Gutcke2021Model}). This will allow us to quantify the robustness of our predictions to IMF-sampling uncertainties, while allowing for a more robust comparison with spectroscopic observations focusing on limited numbers of individual bright stars.

\section{Discussion and conclusions} \label{CONCLUSIONS}
We have studied how a varying IMF with a metallicity-dependent slope affects the observable properties, such as mass and metallicity, of a faint dwarf galaxy. We performed zoom-in, cosmological simulations capable of resolving individual supernovae explosions ($\mdm = 960 \, \msun$, $3 \, \pc$), re-simulating a $\sim 8 \, \times \, 10^{8} \, \msun$ host dark matter halo with varying physical models. We complemented the EDGE galaxy formation model (\citealt{Agertz2020EDGE}) with a new implementation of a metallicity-dependent IMF as suggested by observational reports (\citealt{Geha2013}), self-consistently modelling the resulting alterations to the energy, momentum, radiation and metal budget within the dwarf as the IMF varies.

Under this framework, the IMF becomes top heavy in stellar populations born with lower gas metallicities. In the metallicity range relevant to UFDs ($ -3 \leq{\feh}\leq-1$), the subsequent increase in the fraction of massive stars per population boosts the overall number of supernovae and UV luminosity (Figure~\ref{fig1}), enhancing the effectiveness of supernova and photoionization feedback at regulating star formation. 

This increase results in a hundred-fold suppression of the final stellar mass of the galaxy at fixed halo mass today compared to a canonical IMF (Figure~\ref{fig2}). Furthermore, because we re-simulate the same halo and growth history every time, the ultra-faint's final size does not significantly vary with each re-simulation. We therefore demonstrate that such IMF variations provide a new mechanism to create very low-surface-brightness, field dwarfs (Figure~\ref{fig3}), which will become observable with the next-generation of wide and deep photometric surveys such as the Vera Rubin Observatory. 

In parallel to boosting the available feedback budget, a top-heavy IMF at low ISM metallicities enhances the production of metals in each stellar population (Figure~\ref{fig1}). The net result is a reduction of a dwarf's stellar mass, while maintaining nearly constant iron content (Figure~\ref{fig4}). Our implementation of IMF variations therefore provides a mechanism to produce faint, yet enriched, UFDs. This allows us to populate the faint-end ($\magv \gtrsim - 6$) of the observed mass--metallicity relation, which has remained a significant challenge for modern, numerical simulations of UFDs (e.g. \citealt{Wheeler2019,Agertz2020EDGE, Applebaum2021}). 

Our study, thus, establishes a metallicity-varying IMF as a plausible and compelling framework to interpret observations of UFDs, strongly motivating further studies. A key line of future work will be to extend our findings to multiple objects and formation histories, in order to quantify whether this change in physical model remains compatible with the overall observed population of faint and higher mass, classical dwarfs. Furthermore, our results reaffirm the faint end of the mass--metallicity relation as an acute probe of galaxy formation models. This sensitivity encourages further model testing in the ultra-faint regime, for example, extending our approach to a physically motivated ansatz for how the IMF varies (e.g. with local ISM density, pressure or turbulence; \citealt{Bonnell2006, Marks2012, Hopkins2013}), alternate IMF shapes (e.g. \citealt{Yan2020}) or different empirically motivated scalings (e.g. with stellar velocity dispersion; \citealt{Cappellari2012}). Undertaking this quantification will be key to derive constraints from current and coming photometric and spectroscopic surveys, particularly in light of the prospect to detect isolated, field UFDs in the near-future. This will offer a promising avenue to distinguish ISM-based alterations, like a varying IMF or additional feedback channels (e.g. \citealt{Agertz2020EDGE, Smith2020PhotoRT}), from environmental processing by a Milky-Way host (e.g. \citealt{Applebaum2021}).

\section*{Acknowledgements}
We thank Tobias Buck and the anonymous referees for their comments that helped clarify this manuscript. MP, MR, EA, ASO, OA and FR acknowledge support from the Knut and Alice Wallenberg Foundation, the Swedish Research Council (grants 2014-5791 and 2019-04659) and the Royal Physiographic Society of Lund. MR is further supported by the Beecroft Fellowship funded by Adrian Beecroft. This project has received funding from the European Union’s Horizon 2020 research and innovation programme under grant agreement No. 818085 GMGalaxies. AP is further supported by the Royal Society. This work was performed in part using the DiRAC Data Intensive service at Leicester, operated by the University of Leicester IT Services, which forms part of the STFC DiRAC HPC Facility (www.dirac.ac.uk). This work was also supported by a grant from the Swiss National Supercomputing Centre (CSCS) under project ID s890. Simulations were performed in part using computational resources at LUNARC, the centre for scientific and technical computing at Lund University, on the Swedish National Infrastructure for Computing (SNIC) allocation 2018/3-649, as well as allocation LU 2018/2-28 thanks to financial support from the Royal Physiographic Society of Lund. The authors acknowledge the use of the UCL Grace High Performance Computing Facility, the UCL Hypatia facility provided by the UCL Cosmoparticle Initiative, the Surrey Eureka supercomputer facility, and their associated support services.

\section*{Data Availability}
The data underlying this article will be shared on reasonable request to the corresponding author.

\bibliographystyle{mnras}
\bibliography{IMF_bib}

\appendix
\section{IMF sampling} \label{app:imfsampling}

Cosmological simulations performed in this study are reaching sufficiently high resolutions to potentially introduce stochasticity due to the discrete sampling of the IMF (e.g. \citealt{Su2018, Grudic2019, Applebaum2020, Smith2021IMF}). In particular, the mass of stellar particles used in this work ($300 \, \msun$) is robust to sampling noise for a canonical IMF (\citealt{Smith2021IMF}), but this robustness remains to be quantified for IMFs with a varying slope. In this Appendix, we show that the IMF integrated over the course of a dwarf's history is well-sampled, ensuring that the overall budget of energy, momentum, metal, and radiation from massive stars is stable against Poisson noise arising from sampling the IMF.

To verify this, we compute the distribution of stars in a given stellar mass bin drawn from a single realization of the IMF in a $10^4 \, \msun$ stellar mass population. This stellar mass is larger than individual birth masses of our star particles, but is representative of the total stellar mass formed over the history of our UFDs (Figure~\ref{fig2}). We show in Figure~\ref{figA1} the obtained medians and 68 per cent confidence intervals for four high-mass-end slopes of the IMF.

The number of stars in each bin above $8 \, \msun$ (our assumed limit for massive stars modelling, dashed line) shows limited noise around the median for slopes between 0.85 and 1.85, which range across the metallicity distribution of our UFDs ($-3.0 \leq \feh \leq -1.0$; Figure~\ref{fig1}) and bracket the available IMF over their star formation history. Quantitatively, the number of massive stars reads $N(m > 8 \, \msun) = 276^{+7}_{-8}$ for $\alpha = 1.35$  (corresponding to $\feh=-2.0$), which translates into a total $\mathrm{SNeII}$
energy budget converged to below 10 per cent ($\esnii = 2.76^{+0.07}_{-0.08} \times 10^{53}$ erg). We, thus, conclude that the total distribution of massive stars is well-sampled over an entire dwarf galaxy's history, ensuring the robustness of trends in observables best interpreted through shifts in the overall, integrated feedback and metal budgets between models (Section~\ref{RESULTS}).

Individual star particles, however, remain subject to sampling noise, which will affect the instantaneous coupling between feedback channels and the local ISM. It remains unclear how this uncertainty translates to the final dwarf's observables -- quantifying the noise expected in our predictions from sources of stochasticity will, thus, be essential for future dedicated studies, whether stochasticity arises from IMF sampling at such high resolutions or through chaotic amplification and random seeding of star formation algorithms (e.g. \citealt{Genel2019, Keller2019, Pontzen2021}).

\begin{figure}
    \centering
    \includegraphics[width=\columnwidth]{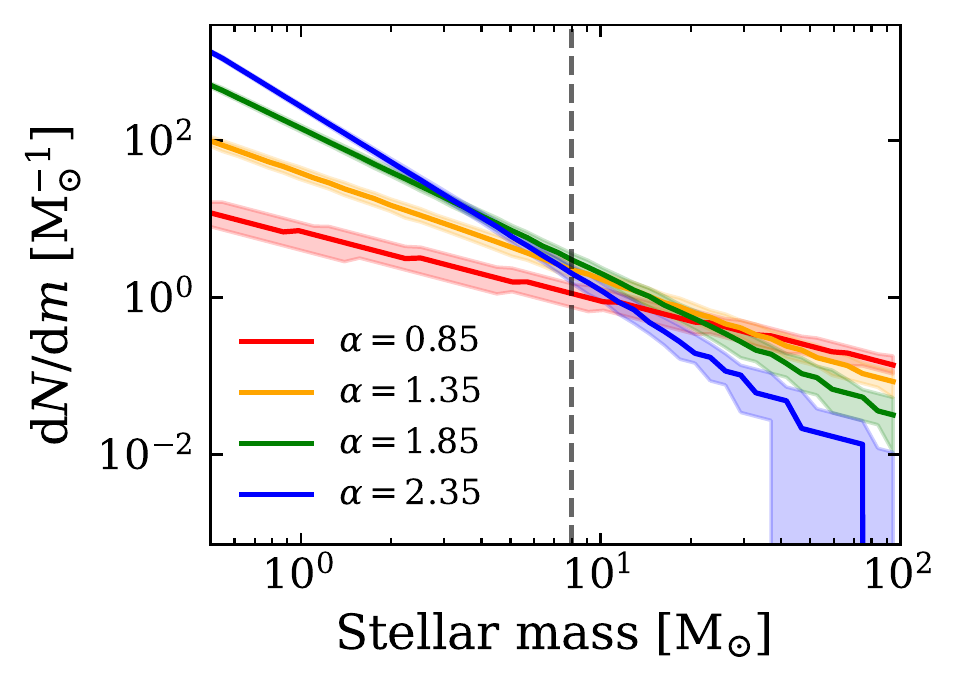}

    \caption{Mass frequency of stars for four high-mass-end slopes of the IMF which bracket our range of variations with metallicity. Lines and contours show the medians and 16--84 scatter in the distribution of stars for a single realization of the IMF over a total mass of $10^4 \, \msun$, representative of the total integrated stellar mass formed within our UFDs (Figure~\ref{fig2}).}
    \label{figA1}
\end{figure}

\bsp 
\label{lastpage}
\end{document}